\documentstyle[buckow]{article}
\begin {document}
\def\titleline{
Magics of M-gravity \hfill January 2001, LPT-ENS/01-03
}
\def\authors{ Bernard L. JULIA
}
\def\addresses{
Laboratoire de Physique Th\' eorique de l'ENS\\
24 rue Lhomond 75231 Paris CEDEX 05 FRANCE.

}
\def\abstracttext{

Compactifications on tori may seem to have revealed their beauty long ago
but the mystery of 11d Supergravity remains and fresh attempts at a conceptual 
breakthrough are worth the effort and quite timely.  
We shall concentrate here on the analogy with Instanton Mathematics and discuss
some open questions and work in progress. 

}
\large
\makefront
\vfill
Talk presented at the Berlin
RTN Workshop, October 2000. Supported by contract HPRN-CT-2000-00131.
\eject
\section{Introduction}
This short presentation covers only one section of the talk and 
has been expanded accordingly. We refer to other conference talks
by the author for other aspects. It has been a fascinating task to 
rewrite the equations of toroidally compactified maximal supergravity as 
self-duality equations similar to the celebrated Instanton equations of four 
dimensional Yang-Mills theory. Let us recall that the latter are first order 
equations obeyed by a subset of the solutions to the full system of classical
equations which are usually written as second order equations in the gauge 
potentials. These equations require  euclidean or (2,2) signature in four 
dimensions. 

In two dimensions analogous 
self-duality equations can also be defined 
for sigma models with target space a Kaehler manifold (N=2),
and if one considers again euclidean signature one needs a 
``twisted'' self-duality equation to compensate for the minus sign in the
square
of the Hodge dualisation. Depending on the signature of spacetime and in 
particular on the dimension there is sometimes the need and always the 
possibility of a twist of the self-duality equations by an invariant operator 
of the symmetry group of square $\pm 1$ and it will appear below.

The typical equations of relativistic physics are second order but they can 
always be rewritten as first order systems. In fact the main result here will 
be to  rewrite the bosonic matter equations of all massless forms of toroidally
compactified 11d Supergravity in first order form so as to become self-duality 
equations.

This was successively realised for the vectors in 4d N=8 SUGRA, for the 
symmetric space sigma models obtained after reduction to two dimensions of all 
SUGRA theories and it generalises independently of supersymmetry there, then 
for the dilaton field of curved space sigma models reduced from three to two 
dimensions (this allows to enlarge the symmetry from the affine algebra to its 
extension by the diffeomorphisms of the circle) and finally to all forms and
all dimensions of M-gravity including odd dimensions of spacetime.

\section{Supersymmetry and Instantons}
 There is a deep connection between self-duality equations and existence of
unbroken  supersymmetry. Let us recall that the so-called BPS condition
started life as a solvable limit of electromagnetic dyon solutions where the 
similarity
between the adjoint  Higgs field and a spacelike extra component of the
Yang-Mill potential becomes exact, there was no explicit
fermion in this picture. It is
the stationary version of the famous instantonic self-duality equation of
pure Euclidean Yang-Mills theory. 

Subsequently
and case by case a suitable supersymmetric extension of each theory
admitting ``self-dual'' solutions was
constructed in which
a Killing spinor ie a covariantly constant spinor can be interpreted as
an  unbroken supersymmetry of the bosonic
background which implies the saturation of the Bogomol'ny bound. The first
analysis of this phenomenon was given in \cite{OW78} in the case of rigid
supersymmetry. So a
bosonic self-duality equation becomes the condition of preservation of some
supersymmetry and stability can be reinterpreted as the property of the
supersymmetry algebra that some bosonic generators are  squares of
fermionic ones. This is not a totally well defined  algorithm as the Killing
spinor equation may involve covariantisation terms beyond the Lorentz
connection. One should now attack the problem head on and analyze the
possibility to embed any  bosonic ``self-duality'' problem
in a larger theory, possibly with more bosonic fields as well, and to decide
 a priori the maximal possible number of supersymmetries that can be realised
once the bosonic  field content
has been decided. In the bosonic case the converse of dimensional reduction has
been first coined group disintegration and then oxidation, we are advocating
now  the study of natural supersymetrisation ie a superoxidation mechanism
in other words the maximal addition to a bosonic theory of fermionic 
dimensions whenever possible. One can ask the same question in the study of 
calibrations and make some progress there.

In a way the next sections address the opposite
problem: 
we are going to show that all the bosonic matter equations of toroidally
compactified 11d SUGRA can be rewritten as self-duality equations of a
generalised but universal type once one doubles the field content. It is a
standard procedure in the analysis of differential systems to introduce
auxiliary variables to render the system first order. The nontrivial
observation maybe is now
that our rather intricate systems are always defined by a finite
dimensional superalgebra (ie $ Z_2$-graded Lie algebra) and have a
universal form. The occurrence of fermionic symmetries is amusing
for bosonic equations but can be understood from the odd character of odd
degree gauge potentials like the three form of 11d SUGRA
\cite{CJLP98}.

We shall call self-duality equation any equation relating some curvatures
$F$ and of the form
\begin{equation}
F = * S F,
\end{equation}
where $S$ is an operator of square plus or minus one that compensates for the
same property of the Hodge duality,  more precisely $S$ exchanges the
generators of the superalgebra associated to gauge potentials and those
associated to their magnetic duals.

\section{Middle dimension}

The prototype example is of course the 4d
Maxwell equations written in terms of
electric and magnetic potentials with dual field strengths. Similarly in
2d the principal sigma model or more generally the
symmetric space sigma models can be rewritten in the above form, at least for
the propagating degrees of freedom. We recall that the typical structure is
that
of a coset space $KG\backslash G$ where $KG$ is the maximal compact
subgroup of $G$. There are two descriptions, first
the gauge fixed one where one
chooses a representative of each coset but the better one restores the
$KG$ gauge
invariance and allows the symmetry under $G$ to become manifest. In the latter
case however the self-duality (in 2d at this stage) involves the components of
the field strength orthogonal to $KG$ only
\begin{equation}
F = (dg.g^{-1})^{\perp}.
\end{equation}
 In fact a harmonic scalar function  and its conjugate form a first order self
dual pair and one can render the $SL(2, R)$ invariance subgroup of the 2d
conformal group  manifest by the same trick.
 
The main example that led to my conviction that self-duality was a general 
feature is 
the case of the 28 vector potentials of 4d $N=8$ SUGRA that cannot form a
representation of the duality symmetry group $G=E_7(7)$ unless one
combines them with the 28 (Hodge) duals. The scalar fields in that theory obey
the equations of the sigma model $KG\backslash G$ again  and the self-duality
equation for the vectors reads in that case
\begin{equation}
g. F = * S g. F,
\end{equation}
where $g$ stands for the 56 dimensional matrix representation of $G$ and
S has to be an invariant operator for $KG=SU(8)$ \cite{CJ79}. This
structure has
been extended to the compactifications of 11d SUGRA on a 3-torus and on
a 5-torus in \cite{CJLP97} and references therein 
for the field strengths of degree half that of the
spacetime volume form.

\section{All forms}
From there it was natural to try an extension to all fields, and we succeeded
for all bosonic forms leaving aside for the time being the graviton and the
fermions. We expect the fermions to transform only under the compact subgroup
$KG$
and under the Lorentz (spin) group. We are now  going to exhibit a vast
generalisation of $G$ or at least of its Borel subgroup. This is quite typical
of broken symmetries in polynomial situations in which the components of some
group element $g$ appear also polynomially in its inverse $g^{-1}$ which occurs
also as we have seen in the equations. The way to permit this is of course
nilpotence
and  this is why the coset spaces appear usually in their Iwasawa
parametrisation. One must restore the local $KG$ invariance to have simple
formulas for the fermionic couplings and for the full action of $G$.  
We refer to \cite{CJLP98} for the compactified cases but we shall
illustrate our general structure wih the 11 dimensional example; the
4-form field strength has a dual that has a non abelian piece. A compact way to
encode the equation of motion  and the
Bianchi identity is to define a supergroup
element and its field strength or curvature by
\begin{equation}
E = exp(A_3 T) exp (A'_6 T')
\end{equation}
\begin{equation}
F = dE E ^{-1}. 
\end{equation}
This is a generalised sigma model
structure, one pair of generators for each form and its
dual,  a theory is then specified by the choice of a  super Lie algebra law. 
The action of the involution  $S$ is simply the exchange of T and T'.
11d SUGRA is defined by the superalgebra
\begin{equation}
\{T,T\}_+ = T'
\end{equation} 
 and its equations of motion are simply $(1)$.

\section{Partial integrability of gravitational theories}
In two dimensions the so-called totally integrable Hamiltonian systems are the 
compatibility conditions for linear (Lax) systems that generate  infinite 
systems of commuting conservation laws and allow a non-local  transformation
to action-angle variables. They usually involve a scattering parameter. The
four
dimensional self-dual Yang-Mills equations do admit such a Lax pair where in 
effect the scattering parameter breaks the 4d covariance by
selecting the 
corresponding anti-self-dual null plane with flat connection. Various 
truncations relate 4d self-dual Yang-Mills equations to most 2d Lorentz 
invariant integrable systems. One such system is the reduction of pure Einstein
gravity from 4  to 2 dimensions on a compact torus by ignoring the torus 
coordinates, the so-called Geroch group acts again nonlocally on solutions or
more precisely on a space of potentials that covers the space of solutions. On 
the other hand there is chaos in some sectors of Einstein theory and the 
coexistence of integrability and chaos is presumably due to the noncompactness
of the symmetry groups involved.\footnote{Indeed negative curvature 
spaces witness chaotic geodesic flows. In this eprint version we comment that 
the Coxeter group  of the overextension \cite{J82} of the 3-dimensional 
duality group controls the BKL chaotic cosmologies \cite{DHJN}. The overextension of
a Dynkin diagram is its affinisation followed by the addition of yet one more vertex
with a single bond to the affine one.} 

An important question is to examine the 
``integrability'' of our twisted
self-dual systems, one problem is of course the lack 
of covariance of the known regular Hamiltonian formulations. 
It should be distinguished from the same question for 4d gravitational 
instantons which has been answered in the affirmative by R.  Ward.
We may also recall that the linearised gravitino equations of N=1 SUGRA have 
Einstein's equations as compatibility condition.

\section{Magic triangle}
We would very much like to understand better the origin of the duality 
symmetries.  At the classical level and in the low energy approximation to 
M-theory one understands some aspects of these exceptional groups for instance
the regularity in the rank r of the group  for instance r is (11-d) in 
the maximal SUGRA 
case (listed as  N=7 below despite the usual denomination N=8  
for a number 4N of supersymmetry charges). It was 
noticed in 1980 that there is a similar regularity at fixed spacetime dimension 
and varying number N of quartet of supersymmetry generators. 
There is an approximate symmetry of the duality groups under the interchange of 
8-N and d-2, for instance the N=7 series of $E_r$ groups which are the split 
real forms are replaced in dimension 3 by other real forms of the same complex
Lie groups namely $E_{N+1}(C)$. 

It turns out \cite{CJLP99} that the Magic is maximal if one gets the 
inspiration to try to 
oxidise curved space three dimensional sigma models for 
split real form noncompact symmetric spaces, in other words for the scalar 
manifolds of toroidally compactified 11d SUGRA. Doing this one abandons in 
general supersymmetry that was related to the other real forms mentioned above
of $E_{N+1}(C)$. However the oxidation triangle one obtains by lifting these 
three dimensional sigma models when possible to higher dimensions 
is completely symmetric under the above symmetry exchanging dimension and SUSY
 number. The reason is not clear yet but is being actively searched for.

In fact these symmetries are low energy symmetries. It is believed that at the 
quantum level the discreteness of the charges in 4 dimensions for instance 
 should break the symmetry down to an arithmetic subgroup. This  
is a generalisation of the breaking of $SL(2,R)$ to $SL(2,Z)$ that occurs for
 instance in type IIB SUGRA in 10 dimensions.

\section{del Pezzo surfaces}

 Finally let us point
out that the series of groups $E_k$ for $k=3, ... ,8$, namely
$A_1\times A_2, A_4, D_5, E_6, E_7, E_8$ is well known in Mathematics, their
Cartan matrices arise as the intersection matrices of the exceptional divisors
on some projective smooth complex surface obtained from the complex projective
plane by blowing up a finite number of suitably placed points. Technically one 
usually requires the singular points to be in general position and the 
``anticanonical sheaf to be ample''.

The cases $k=0,1,2$ are less obvious and the factor R of scaling symmetries 
disappears both in the discrete quantum theory and probably  in the algebraic 
geometry, or at least it has a more subtle interpretation in both cases.  
Nevertheless the IIA and IIB U-duality groups seem to correspond  also to 
two del Pezzo surfaces of degree 8 (private communication of C. Vafa) and 
the projective plane $CP^2$ has degree 9. A detailed discussion can be found 
for instance in \cite{M}. The degree D is in fact the degree of the projective 
surface for instance  the smooth cubic surface of degree 3 has 27 lines and is 
associated to the Cartan matrix of $E_6$, more generally the rank of the group 
is $r=9-D=11-d$ ie $d-2=D$ (one could add by magic symmetry $D=8-N$!).

Work is in progress to understand also  the other columns of the triangle 
($N=1,...,6$) in this way. They follow very similar patterns and do seem to
come
together. One would also like to relax the hypotheses to allow for the affine 
groups corresponding to two dimensions of spacetime and maybe more...
 
As usual with ADE problems the Cartan matrix is a rather universal object but it
is sometimes difficult to relate two instances of the occurrence of simply laced
groups. One example is the Grothendieck-Brieskorn realisation of ADE 
singularities so to speak on the groups themselves, Arnold has been the
champion of a systematic study of other cases. Here we have been enlarging the
Borel 
subgroup of $E_r$ to a rather big Lie superalgebra, 
 its connection with algebraic geometry  remains to be discovered.

\vskip0.5cm
\noindent
{\large \bf Acknowledgements}
\smallskip

\noindent
I am grateful to Mathematics for converging towards M-gravity, 
in particular the general self-dual structure 
applies to analogous physical instances namely Einstein or Einstein-Maxwell 
theories. I benefited from remarks of  A. Beauville, Y. Manin and C. Vafa and
 conversations with  A. Hanani, M. Henneaux and C. Pope.


\begin{thebibliography}{77}

\bibitem{OW78}
D. Olive and E. Witten, {\bf Supersymmetry algebras that include topological
charges} Physics Letters {\bf 78B} (1978) 97.  
\bibitem{CJLP98} E. Cremmer et al., {\bf Dualisation of dualities II} Nuclear
Physics {\bf B535} (1998) 242.
\bibitem{CJ79} E. Cremmer and B. Julia, {\bf The SO(8) supergravity} Nuclear
Physics  {\bf B159} (1979) 141.
\bibitem{CJLP97} E. Cremmer et al.,  {\bf Dualisation of dualities I} Nuclear
Physics {\bf B523} (1998) 73.
\bibitem{J82} B. Julia, {\bf Kac-Moody symmetry of gravitation and supergravity 
theories} Proc. AMS-SIAM Summer Seminar on Applications of Group Theory in 
Physics and Mathematical Physics, Chicago 1982, LPTENS preprint 82/22, 
Lectures in Applied Mathematics, {\bf 21} (1985) 335. 
\bibitem{DHJN} T. Damour et al., {\bf Hyperbolic Kac-Moody Algebras and Chaos in Kaluza-Klein 
Models} to appear in Phys.Lett. B, hep/th/0104231.
\bibitem{CJLP99} E. Cremmer et al., {\bf Higher dimensional origin of d=3 coset 
symmetries} hep-th 9909099.
\bibitem{M} Y. Manin, {\bf Cubic forms} Elsevier, 1974.

\end{thebibliography}
\end{document}